Treatments for pregestational chronic conditions during pregnancy: emulating a target trial with a treatment decision design

Authors: Mollie E. Wood[1], Chase D. Latour[1], Lucia C. Petito[2]

Affiliations:

1. Department of Epidemiology, Gillings School of Global Public Health, University of North Carolina at Chapel Hill, Chapel Hill, NC
2. Division of Biostatistics, Department of Preventive Medicine, Northwestern University Feinberg School of Medicine, Chicago, IL

Corresponding author:

Mollie Wood

Department of Epidemiology

Gillings School of Global Public Health

University of North Carolina at Chapel Hill

135 Dauer Drive

2101 McGavran-Greenberg Hall

CB #7435

Chapel Hill, NC 27599-7435

mwood@unc.edu

## Abstract

Understanding the effects of pharmacotherapies in early pregnancy on pregnancy outcomes is well-understood to be a challenging problem. While many individuals who are trying to become pregnant may decide to discontinue pharmacotherapies for chronic conditions due to uncertainty surrounding possible effects on a developing fetus, others choose to remain on their medications. Still more may newly initiate pharmacotherapy in early pregnancy, to treat newly identified conditions or because barriers to care have lessened. Many current guidelines surrounding pharmacotherapy in pregnancy are based on observational studies that mix these types of users, introducing prevalent user bias. Additionally, conceptualizing last menstrual period or conception as the date around which we anchor our treatment assignment may result in immortal time bias. As a solution to these issues, we suggest emulating a target trial using a treatment decision design, wherein "time zero" is centered around clinical landmarks where treatment decisions may occur, such as the date of preconception counseling or prenatal care initiation. These ideas are illustrated via protocols for two target trials in large administrative databases: antidepressant use for pre-existing depressive disorder and antihypertensive medication use for mild-to-moderate chronic hypertension. Careful consideration of these issues is critical to the identification of the causal effects of early-pregnancy pharmacotherapies on pregnancy outcomes.

**Introduction**

An increasing number of pregnancies are complicated by chronic conditions,[1,2] many of which require management with medications. There is substantial uncertainty about pharmacotherapy during pregnancy,[3] largely due to the challenges inherent in evaluating possible effects on the developing fetus. These challenges include, but are not limited to, (1) selection bias due to incomplete ascertainment of early pregnancy outcomes like spontaneous abortion,[4–6] and (2) time-related biases, particularly immortal time bias, due to varying lengths of gestation and time-specific etiologically relevant periods.[7,8] Because pregnant people are routinely excluded from randomized trials used to evaluate medication safety and efficacy, clinical guidelines are almost entirely based on data from observational studies. Historically, these studies mix types of users in the "treated" category: long-term or prevalent users of a medication who may continue, discontinue, or switch their pharmacotherapy before or during pregnancy, and new users who initiate treatment in the perinatal period.

Here, we propose that perinatal pharmacoepidemiology studies focused on the treatment of chronic conditions in pregnancy should critically assess whether it is appropriate to combine new and prevalent users in one study. We then illustrate how potential biases arising from the combination of new and prevalent users is exacerbated when we use the estimated date of last menstrual period (LMP) or conception as the anchor for eligibility, treatment assignment, and start of follow up (i.e., "time zero"). To address these challenges, we propose emulating a target trial[9] using a treatment decision design.[10] These ideas are grounded in real-world examples of (1) antidepressant treatment for pre-existing depressive disorder and (2) antihypertensive treatment for mild-to-moderate chronic hypertension in pregnancy.

**Methodological Issues Arising in Perinatal Pharmacoepidemiology Studies**

Conducting rigorous studies of drug safety or effectiveness in pregnancy is challenging, particularly because there are so many opportunities for selection out of pregnancy cohorts.[6] However, we focus here on two sources that are particularly problematic when studying treatment of pre-existing chronic conditions early in pregnancy: bias due to the inclusion of prevalent users, and choice of time zero.

*Combining Prevalent and New Users of Pharmacotherapies*

Pharmacoepidemiology studies that include prevalent users of a medication will tend to under-represent individuals who experienced a side effect or otherwise did not have the desired therapeutic response, while over-representing those for whom the medication was effective,[11] leading to overestimation of treatment benefit and underestimation of harm (i.e., prevalent user bias). The new user design was proposed to limit these biases.[12] In this approach, investigators assemble a cohort of treatment-naïve individuals, and only initiators of treatment are assigned as "treated".[12–15] The new user design shares features with target trial emulation, a novel approach to designing and analyzing observational data that requires investigators to describe and emulate a hypothetical randomized study.[9] Both the new user design and target trial emulation emphasize making study design and analytic choices that minimize selection and immortal time biases, such as aligning eligibility criteria, treatment initiation, and start of follow up at time zero.[16]

Despite these advantages, new user designs are rarely employed in studies of medication safety in pregnancy, for several reasons. First, initiation of a new medication to treat a chronic condition is relatively rare in early pregnancy, limiting the availability of data to address these questions even in large data sources such as electronic health records or insurance claims databases. Second, physicians and patients may be more concerned with the decision to discontinue, switch, continue, or otherwise modify existing pre-pregnancy medication regimes in individuals who are newly pregnant, a question that is not necessarily addressed via a new-user design, constituting misalignment of the estimand and the study design. Finally, some researchers argue that the fetus and the pregnancy are new users,[17] even if the pregnant person has been a long-term user of the medication in question.

However, combining these two patient populations routinely presents methodologic challenges in drug safety/effectiveness studies for chronic conditions in pregnancy. Arguing that the fetus and pregnancy are new users requires an assumption that there are no unmeasured common causes of treatment decisions (continuation of medication) and the study outcome,[17] which is often violated in studies focused on chronic conditions in pregnancy. For example, when evaluating the effect of antipsychotic treatment on risk of gestational diabetes, prevalent users of antipsychotics may be at lower risk for metabolic outcomes, because individuals who had a metabolic adverse event at initiation of treatment before pregnancy are more likely to have discontinued and are no longer included among the prevalent users in the pregnancy study (**Figure 1**).[17] Failure to account for this source of bias in studies of pharmacologic treatment for chronic conditions in pregnancy ultimately limits the utility of these important treatment effect estimates.

*Choice of Time Zero*

Another potential source of bias is failure to align treatment assignment with study eligibility criteria and the beginning of follow-up. Misalignment of study baseline, or "time zero," may result in bias due to the presence of immortal time and/or selection on post-exposure variables.[16] Recent work on target trial emulation in pregnancy has focused on choosing the appropriate time zero for the research question, particularly with respect to gestational timing of treatment relative to the outcome of interest. For example, Hernandez-Diaz *et al.* recommend that studies on the risk of birth defects associated with medication use should specify a periconceptional target trial and potentially anchor time zero at the date of the last menstrual period (LMP) or estimated date of conception, while studies on the risk of preterm delivery could consider specifying time zero as any time during pregnancy until the end of gestational week 37,[18] at which point a pregnancy is no longer at risk for preterm delivery. Caniglia *et al.*, in a study of the effect of antibiotic use on preterm delivery, describe a sequence of gestational-week specific trials beginning at week 20 and continuing until week 36, with the aim of reducing the impact of immortal time bias.[19] Schnitzer *et al.* additionally consider gestational age specific treatment effects for scenarios in which treatment might be interrupted by delivery,[20] and Chiu *et al.* suggests a framework for estimating preconception treatments in the presence of competing events.[21] The proposed approaches are substantial methodologic improvements over previous work, but may have important limitations for questions related to treatment for pregestational chronic conditions, and especially for questions about treatment effects in early pregnancy.

For illustration, **Figure 2** shows a study design schematic for a study of medication use in pregnancy. It specifies a range of anchors that researchers might consider as candidates for time zero, including LMP, conception, delivery or end of pregnancy, and time periods corresponding to trimesters or limits of viability. While these anchors certainly have biological significance, there is no guarantee that they correspond to clinical encounters or even recognition of a pregnancy. Other anchors, such as the planning or recognition of pregnancy may be difficult to measure, particularly in large administrative databases. Still other anchors representing occasions when changes to treatment are most likely to occur, such as initiation of prenatal care, are likely to be reliably recorded, but are largely treated as covariates if they are considered at all. These latter anchors related to clinical encounters may be a missed opportunity to reduce bias due to immortal time and selection.

**Identification of causal effects of medication use in early pregnancy**

**Figure 3** visualizes possible underlying causal models where the goal is to estimate the average outcome $Y$ under interventions on $A$, where $A$ represents a time-varying treatment occurring early ($A_0$) and later ($A_1$) in pregnancy and $Y$ an outcome occurring at the end of pregnancy after $A_1$. $S$ is a selection event, such as an induced or spontaneous abortion. To reflect that outcomes are only observed among those who are selected, $S$ is conditioned on and arrows emanating from $S$ are depicted as dashed lines.[22] For simplicity, we have omitted other nodes. Primarily, we are interested in 3 types of causal effects: the average outcome after an intervention on early treatment $\mathrm{E}[Y^{a_0} - Y^{a_0'}]$, the average effect after an intervention on later treatment $\mathrm{E}[Y^{a_1} - Y^{a_1'}]$, and the average effect under a joint intervention $\mathrm{E}[Y^{a_0 a_1} - Y^{a_0' a_1'}]$.

In **Figure 3A**, early treatment $A_0$ is randomly assigned and affects later treatment $A_1$ and $U$ is an unmeasured common cause of both selection and the outcome.

Treatment at both times affects the outcome, Y, but not selection, S. Here, all 3 types of effects are identifiable, as conditioning on S blocks possible backdoor paths through U. In **Figure 3B**, $A_0$ is still randomized, but we introduce an additional complication: earlier treatment $A_0$ affects selection S. Here, the only identifiable effect is of an intervention on early treatment $A_0$. As S is conditioned on, a backdoor path is opened from $A_0$ to U through S, preventing identification of the effect of an intervention on later treatment $A_1$ and a joint intervention on $A_0$ and $A_1$. These effects may be identifiable if methods for dealing with competing risks are implemented.[21] In **Figure 3C**, U is a common cause of early and later treatment as well as selection and the outcome. Here, a further assumption that all U are measured is needed for identification of all 3 effects.[23]

Notably, there are differences between the target population and the observed data, particularly when conducting pregnancy studies in routinely collected health care data. The observed data differs from the data generating model in important ways, namely that the latter includes all pregnancies, while the observed data include only pregnancies that progressed to a certain point. More succinctly: pregnancies do not enter the cohort at LMP, but at the first healthcare encounter at which the pregnancy was registered or known. The observed data are inherently selected: only pregnancies that survive to a certain gestational age and are recorded in the healthcare system are present. This opens a backdoor path for any common causes of S and Y, measured or unmeasured (included in the DAG as U). While it is theoretically possible to identify and measure all U, this is unlikely to be a plausible assumption. Further, early selection events, including those preceded by early treatment, are unlikely to appear in observed data. Thus, estimated effects of early exposure occurring before a pregnancy is recorded in the healthcare system are likely to be biased, and the magnitude and direction of said bias will be difficult to predict, even when treatment is randomly assigned.

**Specifying Time Zero: Potential Candidates**

The majority of pregnancy studies using routinely collected health care data identify pregnancies based on having observed the outcome, making the true time of eligibility the end of pregnancy.[24,25] The misalignment between the target population and the observed, selected analytic sample means that any pregnancy time before contact with the healthcare system is effectively immortal. Timing of pregnancy recognition and prenatal care initiation is highly variable, but recent population-based estimates suggest that in the US, about 15% of pregnant people initiated prenatal care after the 12th week of gestation, and 11% received no prenatal care at all.[26] In routinely collected data, pregnancies are theoretically eligible at LMP or on the date of conception, but truly eligible on the date that the pregnancy is recorded. **Figure 4A**. shows the ideal scenario, in which all pregnancies are enrolled on the date of the last menstrual period, at which point treatment is assigned and follow-up begins. **Figure 4B** shows a more common scenario in which pregnancies are eligible for inclusion in the study if their outcome is observed. Here, potential immortal time could range from being minor or ignorable for pregnancies recognized and registered early, to, effectively, the entire pregnancy, for pregnancies that were unrecognized, did not undergo prenatal care, or received care outside the healthcare system providing the data for study. This is consistent with previous work noting a possible emulation failure as scenario in which start of follow up and treatment assignment occur before eligibility criteria are met.[16] **Figure 4C** is a variation on **4B** in which pregnancies are identified prospectively, becoming eligible at first contact with the health care system related to the pregnancy. Both **4B** and **4C** evaluate exposures occurring from LMP onwards, leaving them vulnerable to bias from immortal time. By contrast **4D** aligns eligibility, start of follow-up, and treatment assignment at the first prenatal care visit. Critically, the time between LMP and prenatal care initiation does not contribute to follow-up time, and any indicators of medication use may be considered as part of the baseline history for an individual pregnancy, but effects of these very early exposures are not estimated. **4E** shows a possible variation that sets time zero at a preconception counseling visit, i.e., the time a treatment decision might be made among individuals planning a pregnancy.

There are compelling reasons for using all available exposure time from LMP onwards, as in scenarios **4B** and **4C**. Early pregnancy is the etiologically relevant window for exposures causing birth defects and is likely also important for early pregnancy events such as placentation that will have profound effects on later pregnancy outcomes. However, the very factors that make this time period so important are what make it difficult to study: the fragility/sensitivity of very early pregnancy means that early losses are common, and many of these losses will go unrecorded in routinely collected healthcare data.

**Emulation of a Target Trial with a Treatment Decision Design**

Randomized controlled trials (RCTs) are the accepted benchmarks for target trial analyses but are especially rare in pregnancy. In **Tables 1** and **2,** we summarize protocols for target trials based on two pragmatic RCTs: the "Stop or Go?" and the Chronic Hypertension and Pregnancy (CHAP) trials. The "Stop or Go?" study recruited pregnant people with current selective serotonin reuptake inhibitor (SSRI) use for a depressive disorder that was in remission or recovered. Patients were then randomized to continue their SSRI treatment with usual prenatal care or to discontinue their SSRI and, instead, attend weekly cognitive therapy sessions.[27] The CHAP study recruited pregnant people with mild-to-moderate chronic hypertension that was known prior to pregnancy or diagnosed for the first time early in pregnancy.[28] These patients were then randomized to receive antihypertensive pharmacotherapy for mild-to-moderate chronic hypertension or to withhold antihypertensive pharmacotherapy unless their hypertension became severe. Both trials were pragmatic, with patients and providers unblinded to treatment assignment.

We have made several important alternations to the proposed target trial protocols to make their emulation in administrative healthcare data more tractable. One major change in the CHAP study target trial is in the "Assignment Procedures:" we have included stratified randomization by antihypertensive use prior to pregnancy. This stems from our argument above: we expect that combining new and prevalent users in this emulation study would result in prevalent user bias; these subpopulations will be analyzed separately. In the "Stop or Go?" target trial, we have altered the "Treatment Strategies" to allow a less specific treatment tapering protocol, as capturing the detailed titration guidance given in the trial is not possible in administrative data, but we could expect to capture discontinuation by 4 weeks after treatment assignment. Further, we have anchored each target trial so that the "Follow-up Period" starts at a prenatal visit where the inclusion criteria are satisfied and at which treatment assignment occurs. We follow this same approach in our suggested emulation: follow-up begins at the first prenatal care visit in the administrative data that satisfies inclusion criteria. This comparison highlights a distinct point about target trial emulation in pregnancy: real trials recruit and randomize people in early pregnancy, not at LMP or conception. It is critical to consider study designs that can emulate this reality (**Figure 4**).

The approach of anchoring time zero of our target trial at a prenatal care encounter is based on the treatment decision design, which is a variation on the new user design.[10] Unlike the new user design, the treatment decision design does not require assembling a cohort of non-users and anchoring treatment assignment upon initiation of a medication. Instead, the treatment decision design is more flexible, anchoring treatment group assigned upon a time point when a decision would be made, such as following the results of a lab test.[10] Studying medication safety and effectiveness for chronic conditions in pregnancy is particularly well-suited to this design, as there are multiple clinically meaningful treatment decision points. For example, healthcare encounters occurring during the periconceptional period could be treated as potential treatment anchors, as they are times when individuals using a medication are either planning to become pregnant or have recently realized they are pregnant, and thus must decide their preferred/best treatment plan for a chronic condition. Similarly, the first prenatal care visit in the first trimester might be appropriate. Importantly, explicitly anchoring studies on healthcare encounters provides some assurance that later selection events would be captured, while still allowing for consideration of the specific gestational timing of the treatment.

**Discussion**

This paper provides an initial framework for studies of medications used to treat chronic conditions in pregnancy, using target trial emulation with the treatment decision design. We propose using contacts with the healthcare system, such as preconception counseling, pregnancy test results, and the first prenatal care visit as the time zero for analysis. We also demonstrate how this can be accomplished using the target trial framework, with an example emulating a recent pragmatic randomized trial for antihypertensive treatment during pregnancy. The complex issue of selection bias, collider stratification bias, and bias arising from time-fixed exposure definitions is well known in the perinatal literature.[6,8,29–32] The current work has many parallels to previous investigations on live birth bias.[5,33] However, to our knowledge, the question of how to include prevalent users of treatments for chronic illness is unresolved, and additional analytic guidance in this topic is needed.

Studies of medication safety in pregnancy are particularly vulnerable to biases caused by conditioning on future events, which might include pregnancy outcomes such as non-live birth, preterm delivery or small for gestational age, as well as outcomes occurring into the childhood of the offspring. It has not escaped our attention that application of this study design may limit our ability to estimate effects of some exposures or in some populations. Studies anchored on initiation of prenatal care may not allow us to estimate the effects of pre-pregnancy medication regime changes and very early pregnancy exposures, particularly in the data sources often used to conduct these studies, such as insurance claims databases and population-based registries.[34] Indeed, it seems clear that some such research questions may need to be framed in terms of a pregnancy-planning or pre-pregnancy cohort, or at minimum, limitations of inferences drawn from pregnancy-only cohorts must be better described. Studies that aim to provide guidance on treatment decisions for use of medications in pregnancy should ensure that the full population making the treatment decision is present in the study sample. It is also important to note that pregnant people who do not initiate prenatal care until late in pregnancy, or whose access to prenatal care is limited due to economic or social reasons, will be systematically excluded from analyses using an early-pregnancy or pre-pregnancy anchor to anchor the treatment decision in question. The impact of these exclusions is likely to vary by region, and particularly by availability and accessibility of affordable health care. Careful attention must be paid to the characteristics of patients included or excluded using different study designs.

Our hypothetical trial might also include as an eligibility criterion that individuals must have initiated prenatal care by a specific gestational period, e.g., the end of the first trimester, with late entries to prenatal care being ineligible for the study. If, for example, a study includes pregnant individuals who enroll after gestational week 17,[35] researchers should be very cautious about interpreting the effects of exposure prior to enrollment particularly if said exposure may have affected the patient's choice or ability to continue their pregnancy. By definition, individuals enrolling in the study had already experienced their observed exposure *and* had a pregnancy surviving beyond the first trimester, making the time before cohort entry effectively immortal.

A trial emulation approach reminds us to evaluate the effect of a treatment decision based only on the information available at the time the decision is made. For example: a pregnant person entering prenatal care at the 9th week of gestation does not know if they will experience a pregnancy loss or preterm delivery, but pregnancy studies frequently limit study samples to live births and/or condition on gestational age at delivery. More subtly, a pregnant person in the 9th week of gestation has already made some decisions about treatment in the first two months of pregnancy. These decisions are fixed by the time of entry to prenatal care, and by definition have not caused a pregnancy loss.

The treatment decision trial emulation anchors the analysis on contact with the healthcare system. In pregnancy, and particularly for research questions concerning medication use in early pregnancy, we expect the most relevant contact for treatment decisions to be the first prenatal care encounter, but other or subsequent clinical encounters may also be of interest, depending on the research question. Critically, this approach allows for the articulation of an intervention that could be carried out and evaluated in the real world, and has clear implications for practice and policy, which is not the case for treatment definitions based on trimesters or similar anchors.

*Conclusion*

Here, we highlighted critical flaws present in many observational studies of the safety and effectiveness of medications used to treat chronic illnesses in pregnancy. We recommended that researchers evaluating the effects of treatments for chronic conditions that may have been present prior to conception implement a treatment decision design anchored on decision points in clinical care to avoid common issues, such as selection bias, prevalent user bias, and immortal time bias, that can strongly bias estimates. Future work is needed to clarify best analytic practices for treatment effects very early in pregnancy.

**Table 1.** Protocol for a target trial based on the "Stop or go?" trial, and its emulation in administrative healthcare data.

| Protocol Component | Description of Target Trial | Description of Emulation in Administrative Healthcare Data |
|---|---|---|
| Eligibility Criteria | • Pregnant individuals between $5^0$-$15^6$ weeks' gestation with a singleton pregnancy<br>• Current use of exactly one selective serotonin reuptake inhibitor (SSRI) for depressive disorder that is in remission or recovered<br>• No history of mania or hypomania, bipolar disorder, suicidality, serious self-harm, any psychotic disorder, current alcohol or drug misuse, or other predominant anxiety or personality disorders requiring psychotherapy >2 times/month<br>• No current severe medical condition | Same, except conditions will be assessed via structured data (e.g., diagnosis codes). |
| Treatment Strategies | 1) STOP: Discontinue SSRIs via a 4-week tapering off protocol under guidance of a psychiatrist. Attend weekly preventive cognitive therapy sessions (≥8 weekly sessions).<br>2) GO: Continue SSRIs. Patients will continue their SSRIs and receive their usual care, which will be monitored. | Same, except pharmaco-treatment will be assessed via structured prescription data (orders and/or fills). |
| Assignment Procedures | Random assignment to exactly one treatment strategy. Randomization (equal allocation) will be stratified by number of prior depressive episodes. Patients and providers are aware of treatment assignment. | We will assume that the two treatment arms will be exchangeable after adjustment for *a priori* identified baseline confounders. |
| Follow-up Period | Follow-up will begin at a prenatal care visit between $5^0$-$15^6$ weeks' gestation, when treatment is assigned, and continue through the earliest of 12 weeks post-partum, | Same, except follow-up will begin at the *first* prenatal care visit between $5^0$-$15^6$ weeks' gestation. |
| | pregnancy loss, or loss to follow-up (60 days without contact with healthcare system). | |
| Outcome | Maternal depressive episode relapse or recurrence during pregnancy through 12 weeks post-partum (binary). | Same, except conditions will be assessed via structured data (e.g., diagnosis codes). |
| Causal Contrast | Intention-to-treat effect; per-protocol effect. | Per-protocol effect, including observational-analogue of the intention-to-treat effect. |
| Analysis Plan | Intention-to-treat analysis; per-protocol analysis: inverse probability weighted (for baseline covariates) logistic regression model. Incorporate censoring at deviation from protocol. Weights estimated as a function of baseline and time-varying post-baseline covariates. Pregnancy loss will be treated as a competing event.[21] | Same, except per-protocol analysis will be conducted in an expanded data set that includes 2 replicates (1 per treatment strategy) for each eligible individual per the clone-censor-weight approach.[9,36,37] |

**Table 2.** Protocol for a target trial based on the "CHAP" trial, and its emulation in administrative healthcare data.

| Protocol Component | Description of Target Trial | Description of Emulation in Administrative Healthcare Data |
|---|---|---|
| Eligibility Criteria | <ul><li>Pregnant individuals prior to $23^0$ weeks' gestation with a non-anomalous singleton pregnancy detected prior to $14^0$ weeks' gestation</li><li>Diagnosis of chronic hypertension prior to pregnancy or $<20^0$ weeks' gestation without preeclampsia or gestational hypertension</li><li>No severe hypertension, secondary cause of chronic hypertension, high-risk comorbidities, ruptured membranes, intrauterine growth restriction, contraindication to commonly used antihypertensives (nifedipine, labetalol).</li></ul> | Same, except conditions will be assessed via structured data (e.g., diagnosis codes). |
| Treatment Strategies | *For users of antihypertensives prior to pregnancy:*<br>1) Continue antihypertensive monotherapy.<br>2) Discontinue antihypertensive monotherapy. Re-initiate antihypertensive medication if hypertension becomes severe.<br>*For non-users of antihypertensives prior to pregnancy:*<br>1) Initiate any* antihypertensive monotherapy.<br>2) Withhold antihypertensive pharmacotherapy unless hypertension becomes severe. | Same, except pharmaco-treatment will be assessed via structured prescription data (orders and/or fills).<br><br>All medication changes are assumed to be due to side effects. |
| Assignment Procedures | Random assignment to exactly one treatment strategy, stratified by antihypertensive pharmacotherapy use prior to pregnancy. Patients and providers are aware of treatment assignment. | We will assume that the two treatment arms will be exchangeable after adjustment for *a priori* identified baseline confounders. |
| Follow-up Period | Follow-up will begin at a prenatal care visit between $14^0$ and $22^6$ weeks' gestation and continue through the earliest of 6 weeks post-delivery or loss to follow-up (60 days without contact with healthcare system). | Same, except follow-up will begin at the *first* prenatal care visit between $14^0$ and $22^6$ weeks' gestation. |
| Outcomes | Composite of preeclampsia with severe features ≤2 weeks after delivery), medically indicated preterm birth (<35 weeks' gestation), placental abruption, or fetal or neonatal death (≤28 days postpartum). | Same, except conditions will be assessed via structured data (e.g., diagnosis codes). |
| Causal Contrast | Intention-to-treat effect; per-protocol effect. | Per-protocol effect, including observational analog of intention-to-treat effect |
| Analysis Plan | Intention-to-treat analysis; per-protocol analysis: inverse probability weighted (for baseline covariates) logistic regression model. Incorporate censoring at deviation from protocol. Weights estimated as a function of baseline and time-varying post-baseline covariates. All analyses will be stratified by use of antihypertensives prior to pregnancy. | Same, except per-protocol analysis will be conducted in an expanded data set that includes 2 replicates (1 per treatment strategy) for each eligible individual per the clone-censor-weight method. [9,36,37] |

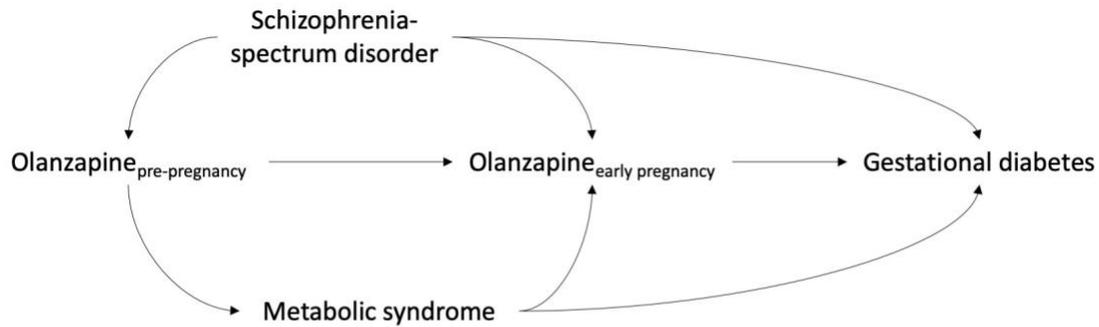

*Example from* Huybrechts KF, Bateman BT, Hernández-Díaz S. Use of real-world evidence from healthcare utilization data to evaluate drug safety during pregnancy. *Pharmacoepidemiol Drug Saf*. 2019;28(7):906-922.

**Figure 1.** Directed acyclic graph (DAG) demonstrating that inclusion of prevalent users of second generation antipsychotics such as olanzapine may result in an underestimation of the effect of exposure on gestational diabetes, as prevalent users are at lower risk for metabolic syndrome, assuming that users susceptible to metabolic effects of the medication will have discontinued or switched their medication.

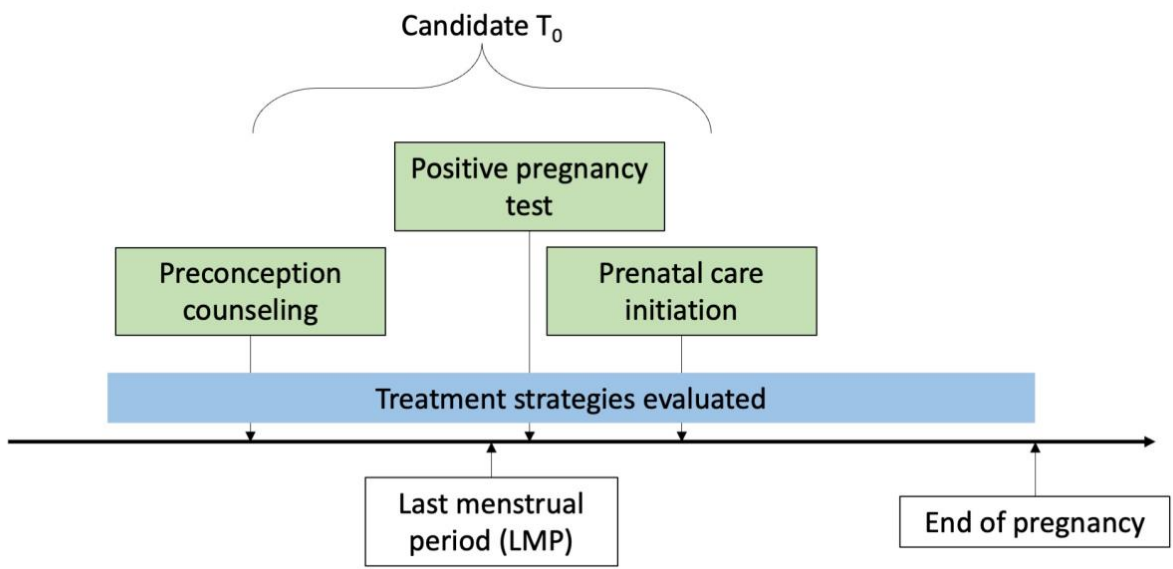

**Figure 2.** Study design schematic noting typical landmarks used in studies of medication safety in pregnancy (LMP, end of pregnancy) and potential alternate landmarks tied to encounters with the healthcare system (preconception counseling, positive pregnancy test, prenatal care initiation). Time zero ($T_0$) could plausibly be tied to any of these landmarks, depending on the research question.

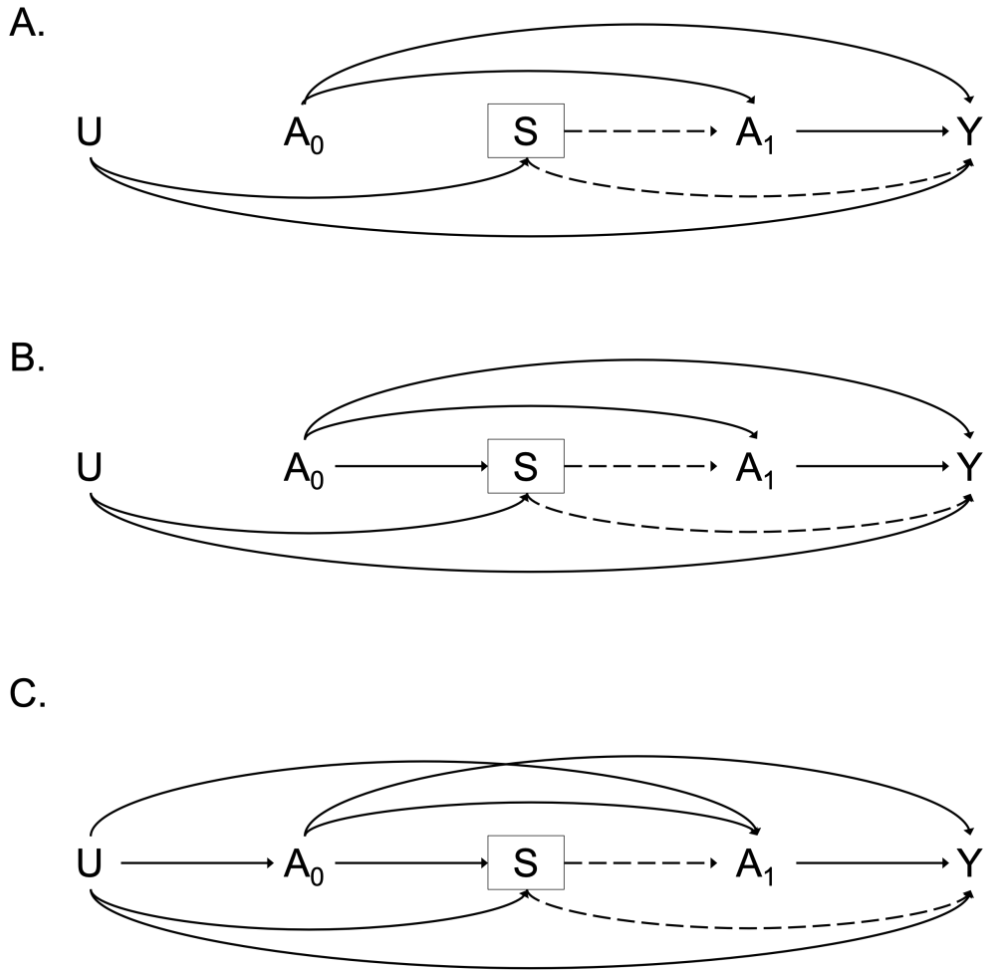

**Figure 3.** Three directed acyclic graphs depicting possible data generating mechanisms. Panel A represents a scenario with randomization of treatment in early pregnancy, later treatment $A_1$ affected by earlier treatment, and independence between $A_0$ and selection. Panel B represents a scenario in which early treatment additionally affects selection and future treatment. Panel C depicts a scenario in which early treatment is no longer randomized, and the set of unmeasured common causes of selection and the outcome also affect early and later treatment. $A$ is a time varying treatment measured early ($A_0$) and later ($A_1$) in pregnancy. $S$ is a selection event, such as an induced or spontaneous abortion, and $Y$ is an outcome occurring at the end of pregnancy. $U$ indicates unmeasured common causes of $S$ and $Y$ (panels A and B) and treatment $A_0$ and $A_1$ (panel C). To indicate the dependence induced by the selection event $S$ (e.g., the outcome $Y$ by definition does not occur in the presence of a competing event $S$), arrows emanating from $S$ are depicted as dashed lines.

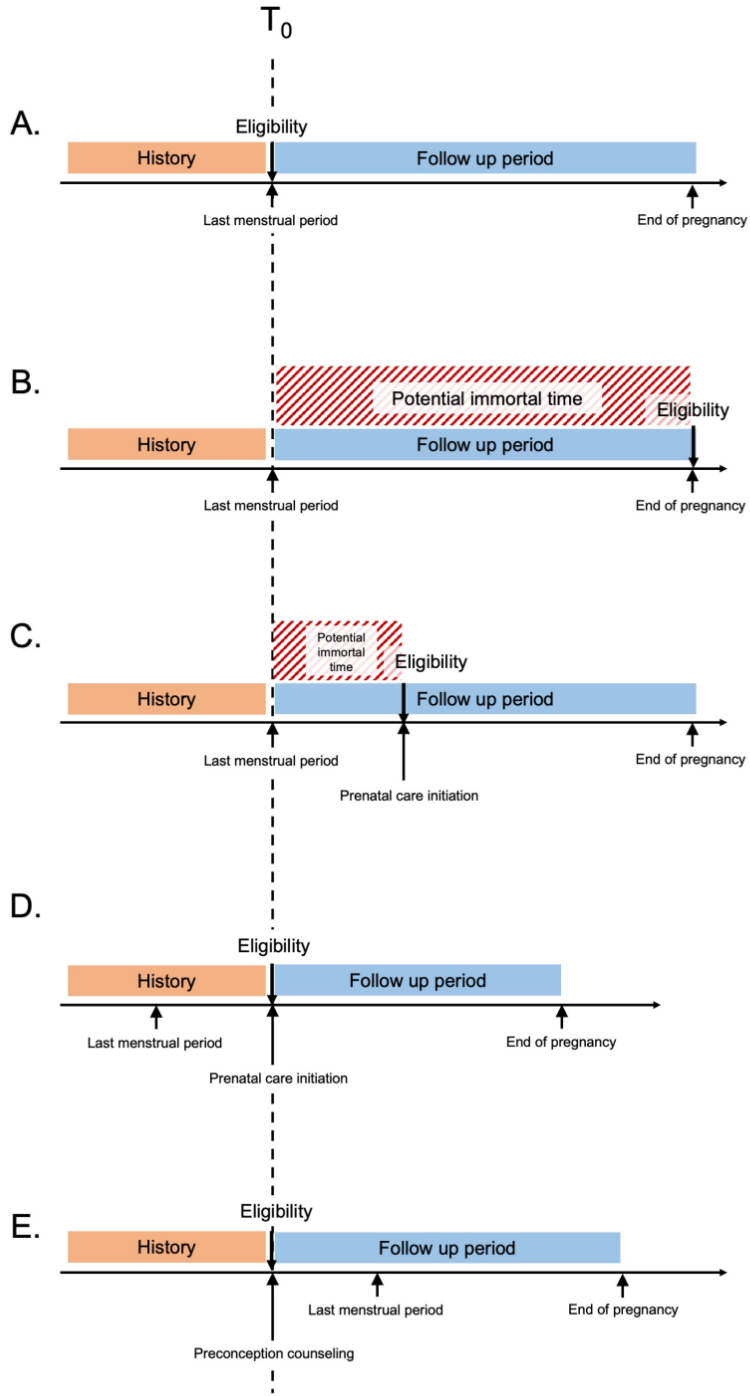

**Figure 4.** Five scenarios representing timing of eligibility, treatment assignment, and start of follow-up at two commonly used pregnancy landmarks (date of last menstrual period, end of pregnancy) compared with standard clinical contacts (prenatal care initiation, preconception counseling) with implications for bias arising from misaligned time zero ($T_0$). Panel A represents the ideal scenario with no self-inflicted bias, where all individuals are enrolled on the date of the last menstrual period, treatment is assigned,

and follow-up begins. Panel B represents a common scenario where outcome observation, occurring as late as end of pregnancy, is required for study inclusion; all time between $T_0$ and end of pregnancy can potentially be immortal. Panel C represents the scenario where pregnancies are identified prospectively, but eligibility is only possible at first contact with the health care system. Immortal time here is still present but limited to time between last menstrual period (and assumed treatment assignment) and prenatal care initiation. Panels D and E represent scenarios where eligibility, start of follow-up, and treatment assignment are aligned at either the first prenatal care visit (D) or a preconception counseling visit (E).